\documentclass[notitlepage,twocolumn,prl,nobalancelastpage,amssymb,superscriptaddress,showpacs,longbibliography]{revtex4-1}
\usepackage{graphicx}
\usepackage{amsmath}
\usepackage{siunitx}
\usepackage{amssymb}
\usepackage[colorlinks=true,citecolor=blue,linkcolor=blue]{hyperref}
\usepackage{float}
\usepackage{lipsum}
\usepackage{comment}
\usepackage{cancel}
\usepackage{soul}
\usepackage{tikz}
\usepackage[dvipsnames]{xcolor}
\usepackage{physics}
\usepackage{commath}

\def\bea{\begin{equation}\begin{aligned}} \def\eea{\end{aligned}\end{equation}}
\def\gcg{\tilde{g}_{\pm}}

\definecolor{TScolor}{RGB}{255, 153, 51}

\definecolor{CYcolor}{RGB}{50, 100, 50}

\newcommand{\Ch}[1]{\textcolor{black}{#1}} 

\begin{document}

\author{Christopher Yang}
\address{Department of Physics and Astronomy, University of California, Irvine, Irvine, California 92697, USA}

\author{Thomas Scaffidi}
\address{Department of Physics and Astronomy, University of California, Irvine, Irvine, California 92697, USA}

\title{Asymptotic freedom, lost: Complex conformal field theory in the two-dimensional $O(N>2)$ nonlinear sigma model and its realization in Heisenberg \Ch{spin chains}}

\begin{abstract}
The two-dimensional $O(N)$ nonlinear sigma model (NLSM) is asymptotically free for $N>2$: it exhibits neither a nontrivial fixed point nor spontaneous symmetry-breaking.
Here we show that a nontrivial fixed point generically does exist in the \emph{complex} coupling plane and is described by a complex conformal field theory (CCFT).
This CCFT fixed point is generic in the sense that it has a single relevant singlet operator, and is thus expected to arise in any non-Hermitian model with $O(N)$ symmetry upon tuning a single complex parameter.
We confirm this prediction numerically by locating the CCFT at $N = 3$ in \Ch{two} non-Hermitian spin-1 antiferromagnetic Heisenberg \Ch{chains, and in a non-Hermitian spin-$1/2$ ladder,} finding good agreement between the complex central charge and scaling dimensions and those obtained by analytic continuation of real fixed points from $N\leq 2$.
We further construct a realistic Lindbladian for a spin-1 chain whose no-click dynamics are governed by the non-Hermitian Hamiltonian realizing the CCFT.
Since the CCFT vacuum is the eigenstate with the smallest decay rate, the system naturally relaxes under dissipative dynamics toward a CFT state, thus providing a route to preparing long-range entangled states through engineered dissipation.

\end{abstract}

\maketitle

\textit{Introduction}.---The $O(N)$ nonlinear $\sigma$-model (NLSM) occupies a central role in quantum field theory and describes a diverse class of systems, such as magnetic systems, superfluids, and chiral effective models in quantum chromodynamics  \cite{Kosterlitz_1973,polyakov1975,brezinZJ1976PRL,bardeenLeeShrock1976,brezinZJ1976PRB,WITTEN1983422,GASSER1984142,friedan1985,gawedzkiKupiainen1986,mitterRamadas1989}.
In two dimensions, it famously flows to strong coupling in the infrared for $N > 2$, the simplest example of asymptotic freedom. This results in the absence of an ordered phase and of a non-trivial fixed point at finite coupling \cite{polyakov1975,bardeenLeeShrock1976,HALDANE1983464,PhysRevLett.61.1029,PhysRevB.93.241108,Fradkin2013,PhysRevB.90.214426}.
These properties reflect the Mermin-Wagner theorem, which forbids the spontaneous breaking of continuous symmetries in 2D \cite{PhysRevLett.17.1133}, and seem to preclude any nontrivial fixed point.

In this work, we show that this conclusion can be dramatically altered in non-Hermitian settings---relevant, for example, to monitored quantum dynamics---where in the field-theory formulation the only change is that the coupling becomes complex, $g\in\mathbb{C}$, in the $O(N)$ NLSM:
\begin{equation}
\label{NLSM}
S \;=\; \frac{1}{2g}\int d^{2}x \, (\nabla \hat{ n})^{2}.
\end{equation}
Here $\hat{ n}$ is an $N$-component unit vector, $\hat{ n}\in S^{N-1}$.

\begin{figure}[t!]
    \centering
    \includegraphics[width=0.6\linewidth]{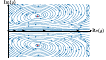}
     \caption{Schematic RG flow (from UV to IR) of the 2D $O(N>2)$ nonlinear $\sigma$-model (NLSM) with a complex coupling parameter $g$. On the real axis, the flow goes to strong coupling. The CCFTs (red dots) appear at complex-conjugate positions in the complex-$g$ plane, around which the flow forms an outward spiral which eventually also flows to strong coupling. }
    \label{fig:intro}
\end{figure}

 As we will show, allowing $g$ to be complex reveals a pair of complex conjugate fixed points described by a complex conformal field theory (CCFT) \cite{Gorbenko2018,10.21468/SciPostPhys.5.5.050,PhysRevLett.124.161601,PhysRevLett.124.051602,Faedo2021,PhysRevLett.131.131601,PhysRevLett.133.077101} with a complex central charge $c(N) \in \mathbb{C}$, e.g. $c \approx 1.51 \pm 0.158i$ for $O(3)$ (see Eq.~(\ref{eq:c}) for the general expression of $c(N)$).
This complex fixed point gives rise to a spiral Renormalization Group (RG) flow in the complex plane for $g$, schematically shown in Fig.~\ref{fig:intro}.
Asymptotic freedom is thus generically lost in the non-Hermitian version of the NLSM 
because there is now a large part of the phase diagram which in the UV 
flows to the non-Gaussian CCFT, rather than the trivial $g=0$ fixed point.

Complex CFTs describe fixed points which have moved to the complex plane through the annihilation of two real fixed points, and have been proposed as a natural explanation for the  conformality loss and the ``walking behavior'' (or weakly first-order transitions) observed on the real axis of a variety of condensed matter and high-energy systems \cite{doi:10.1143/JPSJ.72.74,PhysRevD.80.125005,PhysRevX.5.041048,PhysRevX.7.031051,Gorbenko2018,10.21468/SciPostPhys.5.5.050,doi:10.7566/JPSJ.88.034006,PhysRevB.99.195110,PhysRevB.99.195130,PhysRevLett.124.051602,PhysRevB.102.020407,10.21468/SciPostPhys.15.2.061,PhysRevX.14.021044,PhysRevLett.133.076504,https://doi.org/10.48550/arxiv.2507.14732}. Subsequently, numerical works studied complex fixed points directly by working with non-Hermitian Hamiltonians or transfer matrices in a variety of models, including the $Q>4$ Potts models~\cite{PhysRevLett.44.837,PhysRevB.22.2560,PhysRevLett.63.13,PhysRevLett.133.077101,https://doi.org/10.48550/arxiv.2507.14732,n578-ljd5,PhysRevLett.133.077101} and $O(N>2)$ loop models~\cite{PhysRevLett.131.131601}, which are closely related to the fixed points described here, as we will explain. (See also recent works on quantum spin models~\cite{ZouXuan2025,https://doi.org/10.48550/arxiv.2506.16424}).
However, two critical questions remain: (1) How generic are CCFTs---can they be found in a broad class of systems? and (2) Can they be realized in an experimentally relevant non-Hermitian system?

We address these questions as follows. First, we show that the $O(N)$ complex fixed points introduced here are generic: they possess a single relevant singlet operator with $\Re(\Delta_\epsilon) < 2$, and thus emerge by tuning a single \emph{complex} parameter. They are therefore expected in any non-Hermitian system with $O(N)$ symmetry, for $N>2$.

Second, to address physical realization we propose a simple microscopic model---the non-Hermitian spin-1 Heisenberg chain---which realizes the $O(3)$ CCFT fixed point. We confirm our results non-perturbatively by locating this fixed point via exact diagonalization of the non-Hermitian Hamiltonian. Using operator-state correspondence to map the low-lying spectrum $\epsilon_\mu - i \gamma_\mu$ of this Hamiltonian to CFT operators, we find close agreement with predicted scaling dimensions. 
The eigenstate corresponding to the CCFT vacuum exhibits not only the lowest energy $\epsilon_\mu$ but also the slowest decay rate $\gamma_\mu$. Consequently, time evolution under this Hamiltonian naturally relaxes the system to the CCFT vacuum state. Finally, we provide a microscopic Lindbladian for a spin-1 chain whose no-click dynamics (accessible via continuous monitoring and post-processing, see e.g. Ref.~\cite{PhysRevLett.126.170503}) realize this non-Hermitian Hamiltonian.

From an open quantum dynamics perspective, our work establishes CCFTs as a new universality class to describe criticality in dissipative quantum systems~\cite{PhysRevB.85.184302,PhysRevLett.110.195301,Sieberer_2016,PhysRevLett.125.266803,PhysRevB.105.205125,PhysRevX.13.021026,PhysRevB.107.235153, PhysRevB.108.064312,PhysRevX.13.041042}, and provides a distinct example of how dissipation can induce a unique form of criticality that is otherwise impossible in equilibrium.
The CCFT we identify has a complex central charge and is thus distinct from several critical phases previously studied in non-Hermitian systems. In particular, \Ch{the $O(N>2)$ CCFT cannot simply be obtained by complexifying some parameter in a unitary CFT with the same $O(N>2)$ symmetry, as is the case for e.g. dissipative Tomonaga-Luttinger liquids} \cite{PhysRevB.105.205125}, nor is it equivalent to a non-unitary CFT with negative but real central charge, as has been predicted in a number of non-Hermitian settings \cite{CastroAlvaredo2009,PhysRevLett.119.040601,PhysRevResearch.2.033017,10.21468/SciPostPhys.12.6.194,Dengis_2014,zhou2024}. In the context of monitored dynamics and quantum state preparation \cite{lwrz-jxrr,PhysRevResearch.2.033017,Biella2021,PhysRevA.78.042307,Weimer2010,RevModPhys.97.025004}, our work introduces a route to prepare highly-entangled critical phases with universal properties. The complex fixed point we find has logarithmic entanglement scaling with a universal central charge, contrasting with the non-universal effective central charges previously found to govern some dissipation-induced entanglement transitions  \cite{PhysRevB.106.235149,PhysRevB.103.224210,PhysRevB.108.165126,PhysRevLett.126.170602,PhysRevB.111.064313,PhysRevB.104.184422}.

\begin{figure}[t!]
    \centering
    \includegraphics[width=1\linewidth]{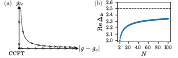}
    \caption{(a) Schematic two-parameter RG flow for the $O(N>2)$ NLSM. The parameter $|g-g_c|$ is the distance from the critical point $g_c$ in the complex plane of the coupling constant $g$, and thus corresponds to the energy operator $\epsilon$, which is relevant.  The parameter $g_{lc}$ controls the strength of the ``loop crossings'' and corresponds to the 4-leg watermelon operator, with scaling dimension $\Delta_{lc} \equiv \Delta_{l=4}$, which is irrelevant at the CCFT. 
Contrary to the loop model, for which $g_{lc}=0$ due to an additional microscopic, non-invertible symmetry, the NLSM generically has $g_{lc}\neq 0$ in the UV. It therefore only flows toward $g_{lc}=0$ in the IR, realizing the same non-invertible symmetry as the loop model, but now as an \emph{emergent} symmetry that appears only at the CCFT fixed point.
    (b) Scaling dimension of loop crossings $\Delta_{lc}$, which are irrelevant for all $N>2$. The dashed line indicates the asymptote $\mathrm{Re}~\Delta_{lc} (N \to \infty) = 2.5$.}
    \label{fig:loopc}
\end{figure}

\textit{$O(N)$ loop model.}---
The $O(N)$ CCFTs we investigate appear naturally in the context of $O(N)$ loop models~\cite{PhysRevLett.49.1062,Baxter_1986,diFrancesco1987,PhysRevLett.61.138,Blote_1989,Jacobsen2009,10.21468/SciPostPhys.12.5.147,10.21468/SciPostPhys.16.4.111}. There, they are the analytic continuation to $N>2$ of two branches of real fixed points which annihilate at $N=2$ and move to the complex plane for $N>2$. 
(In the $O(N)$ loop model, $N$ is a continuous parameter giving the loop fugacity~\cite{PhysRevLett.131.131601}). 
Using Coulomb gas methods, these loop models were shown to have central charge
\bea\label{eq:c}
c_{\pm}(N) = 1 - \frac{6 (1-\gcg)^2}{\gcg} ,
\eea
where $\gcg = 1 \pm e(N)$ and $e(N) = \frac{1}{\pi} \cos^{-1}(N/2)$~\cite{Baxter_1986,diFrancesco1987,PhysRevLett.61.138,Blote_1989,Jacobsen2009,10.21468/SciPostPhys.12.5.147,PhysRevLett.131.131601,10.21468/SciPostPhys.16.4.111}.
For $N<2$, these two branches correspond to two different real CFTs, describing the dense and dilute phases of $O(N)$ loop models (e.g. for $N=1$ the dilute phase describes the Ising CFT).
For $N>2$, $e(N)$ becomes purely imaginary, and the two branches form a complex conjugate pair of CCFTs with complex central charge~\cite{PhysRevLett.131.131601}.

\begin{figure*} [t!]
    \centering
    \includegraphics[width=1\linewidth]{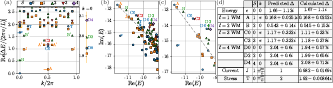}
    \caption{
    {Spectrum of the Heisenberg spin-1 chain at the finite-size complex fixed point [Eq.~(\ref{eq:J2Kvals})] for a length $L = 14$ chain.} (a) Real part of the extracted scaling dimensions, obtained as rescaled energy gaps $\Delta E / (2 \pi v / L)$, where $\Delta E = E - E_g$ and $E_g$ is the ground state energy and $v = 0.796-1.03i$ is the complex velocity. Different colors and symbols indicate states with distinct $O(3)$ total spin number $S$. Letters label states with known $O(N)$ CCFT predictions, as summarized in panel (d). Narrow horizontal lines denote the predicted real parts of the scaling dimensions. States labeled by primed letters correspond to descendants of the primaries labeled by the corresponding unprimed letters. On the right, we zoom in on the orange box enclosing states at $k = 0$ with nearly-marginal scaling dimensions, of which D0, D2, and D4 arise from the $\ell = 4$ watermelon operator. (b) Energy spectrum in the complex plane without rescaling by $2\pi v / L$, with the dashed line parameterized by $E_g + vx$, where $x \in \mathbb{R}^+$ showing the direction of the conformal towers.  Note that we shift the eigenenergies by an imaginary constant $\delta$ (see the discussion following Eq.~(\ref{eq:lindblad})). (c) Same as (b), but only displaying states with momenta $k =0$ and $\pi$. On this panel, the clusters of nearly-degenerate states corresponding to the $\ell=1,2,3,4$ watermelon operators (labeled by A, B, C, D) become obvious. (d) Table of identified CCFT states, comparing the predicted and numerically calculated scaling dimensions $\Delta$. WM stands for watermelon operators.}\label{fig:spectrum}
\end{figure*}

The scaling dimensions also become complex for $N>2$, and can similarly be predicted by analytic continuation of the $O(N \le 2)$ results \cite{10.21468/SciPostPhys.12.5.147,PhysRevLett.131.131601}. In particular, the thermal scaling dimension, associated with the loop tension in the loop model and to the energy operator in the NLSM, is given by  
\begin{equation} \label{eq:xtpred}
    \Delta_\epsilon = {4}/{\gcg} - 2.
\end{equation}

Another set of operators are the $\ell$-leg watermelon operators~\cite{Jacobsen2009}, which in the loop model correspond to having $\ell$ loop strands emanating from or terminating at a point.
They have scaling dimension
\begin{equation} \label{eq:wmpred}
    \Delta_{\ell} = \frac{1}{8} \gcg \ell^2 - \frac{(1 - \gcg)^2}{2 \gcg}.
\end{equation}
For instance, the $\ell = 1$ watermelon operator corresponds to the $O(N)$ vector field $\hat{n}$ and $\Delta_{\ell=1}$ thus governs the power-law decay of the two-point function between $O(N)$ spins \cite{Jacobsen2009}.
 More generally, the multiplicity of watermelon operators for $\ell>1$ appears as the \emph{sum} of distinct $O(\Ch{N})$ irreducible representations, and this was recently explained by the presence of an additional, non-invertible symmetry (associated with topological defect lines)~\cite{Gorbenko2020,10.21468/SciPostPhys.14.5.092,Jacobsen2023} present at the microscopic level in the loop model.

\textit{Predictions for the $O(N)$ NLSM.}--- We now discuss how to port these results from the loop models to the NLSM. What differentiates the two models is that the NLSM allows for ``loop crossings''~\cite{PhysRevLett.90.090601}.
Indeed, $O(N)$ loop models arise as a truncated high-temperature expansion of the $O(N)$ classical Heisenberg model, where the truncation amounts to neglecting loop crossings~\cite{PhysRevLett.49.1062}. 
If loop crossings are relevant, the loop model and the NLSM are in different universality classes.
(This is what happens on the dense branch for $N<2$~\cite{PhysRevLett.90.090601}.)
Conversely, if loop crossings are irrelevant, the loop model fixed points should also describe the generic critical behavior of the NLSM (See Fig.~\ref{fig:loopc}a).
Crucially,  this is what happens for $N>2$: loop crossings, which correspond to the singlet component of the $\ell=4$ watermelon operators~\cite{PhysRevLett.90.090601}, are irrelevant for $N>2$ (see Fig. \ref{fig:loopc}b).
The loop model $O(N)$ CCFT fixed points therefore describe the generic criticality of the non-Hermitian $O(N)$ NLSM as well.
As a side result, it also means that the non-invertible symmetry of loop models introduced in Ref.~\cite{Jacobsen2023} is expected to appear for the NLSM as well, but as an \emph{emergent} symmetry at the CCFT fixed point.

\textit{Microscopic model and numerical results.}--- As a concrete example at $N=3$ with experimental relevance to quasi-1D antiferromagnets~\cite{PhysRevLett.56.371,Renard_1987} and various quantum simulator platforms~\cite{Bloch,PRXQuantum.6.020349}, we study the spin-$1$ Heisenberg chain described by the Hamiltonian~\footnote{We use integer spin $S=1$ so that the $\theta = 2\pi S$ topological term does not change the physics in the bulk.}
\begin{equation}\label{eq:NHH}
    H = \sum_{i} \left[ J_1 \boldsymbol{S}_{i} \cdot \boldsymbol{S}_{i+1} + J_2 \boldsymbol{S}_{i} \cdot \boldsymbol{S}_{i+2} + K (\boldsymbol{S}_i \cdot \boldsymbol{S}_{i+1})^2 \right],
\end{equation}
where $\boldsymbol{S}_i = (S_i^x, S_i^y, S_i^z)$ represents the vector of spin-$1$ matrices $S_i^\alpha$ acting on site $i$, with $\alpha = x, y, z$. 
Here, $J_1$, $J_2$, and $K$ denote the coupling parameters for nearest-neighbor exchange, second neighbor exchange, and nearest-neighbor biquadratic coupling, which are the three simplest local $SU(2)$-symmetric terms, with $J_1$ being typically the largest term (we will take $J_1=1$).

As long as $J_2$ and $K$ are not too large compared to $J_1$, this Hamiltonian realizes a gapped, short-range entangled, paramagnetic phase with symmetry-protected topological order called the Haldane phase.
This phase is well described by the $O(3)$ NLSM of Eq.~(\ref{NLSM}) with a coupling constant $g$ that depends on $J_2$ and $K$, see Refs.~\cite{HALDANE1983464,Affleck:191830,PhysRevB.93.241108,PhysRevB.90.214426}. The flow to strong coupling of the NLSM on the real axis explains the absence of antiferromagnetic order, the finite gap, and the exponentially decaying correlations in the spin-1 Heisenberg chain~\cite{HALDANE1983464}.
By allowing $J_2$ and $K$ to be complex, however, we will 
obtain an NLSM with a complex coupling constant $g(J_2,K) \in \mathbb{C}$ which exhibits a critical fixed point at $g=g_c$ described by a CCFT with power-law correlations.

Before moving on to the numerical results, we address a practical challenge in the numerics. 
As shown in Fig.~\ref{fig:loopc}(a), accessing the CCFTs should require the tuning of a single complex parameter, since there is a single relevant singlet operator $\epsilon$.
However, the loop crossing operator, which is the singlet component of the $\ell=4$ watermelon operator, is only weakly irrelevant for $N=3$: it has scaling dimension $\Re\Delta_{\ell=4} \approx  2.04 $, as shown in Fig. \ref{fig:loopc}(b). 
This means the loop crossing coupling constant $g_{lc}$ decays very weakly with system size as $\sim L^{-0.04}$, which leads to very slow finite size convergence to the CCFT fixed point in our numerics. 
To address this issue, we decide to tune an additional parameter which allows us \Ch{to} minimize the loop crossing strength $g_{lc}$ in the Hamiltonian, thus leading to much faster finite size convergence to the CCFT fixed point.
(That would be less of an issue at larger $N$ since the loop crossing operator becomes more irrelevant as $N$ increases, although it does so slowly, see Fig.~\ref{fig:loopc}(b).)

Using exact diagonalization on a periodic chain of length $L=14$, and a gradient descent method to find the location of a finite-size critical point with optimal finite size convergence, we find a complex conjugate pair of critical points at 
\begin{equation}\label{eq:J2Kvals}
(J_2,K)_{\pm} = (0.0660 \pm 0.338i, 0.176 \pm 0.335i).
\end{equation}
We focus on the point $(J_2,K)_{+}$ in the rest of the Letter, which realizes the $c_+$ branch in Eq.~(\ref{eq:c}).
At that point, the low-energy spectrum, see Fig.~\ref{fig:spectrum}, displays primary and descendant states in close agreement with the CCFT predictions.  Specifically, Fig.~\ref{fig:spectrum}(a) shows the finite-size estimates of the scaling dimensions, calculated as rescaled energy gaps $\Delta E / (2\pi v/L)$, with $\Delta E = E - E_g$ \cite{Cardy_1984}. The ground state, with energy $E_g$, is defined as the eigenvalue with the smallest real part \footnote{Strictly speaking, depending on the value of $\mathrm{Arg}[v]$ another primary could have been the state with the lowest real part of the energy, but this does not happen for the value of $v$ we found.}.
The value of the velocity was obtained numerically $v = 0.796-1.03i$ (see \Ch{the Supplemental Material (SM)~\cite{SupplementalMaterials}}).
As expected for a critical state, the entanglement of the ground state grows logarithmically with subsystem size, and the extracted central charge agrees well with the prediction: $c = 1.529 -0.161i$ within $2\%$ of the predicted value $c_+ = 1.51 - 0.158i$ (see \Ch{the End Matter and SM~\cite{SupplementalMaterials} for details of the central charge calculation and verification of its consistency with the finite-size scaling of the ground-state energy).}

\emph{Identification of CFT operators.}---
Each eigenstate has a well-defined lattice momentum $k$ and $O(3)$ spin quantum number $S$. The first singlet ($S=0$) excited state in the $k=0$ sector is identified with $\epsilon$, the energy operator of the NLSM. The corresponding scaling dimension governs the spiral flow of $g$ around the fixed point, according to $ g -g_c \sim L^{2 - \Delta_\epsilon}$. We find a good agreement between the Coulomb gas prediction $\Delta_\epsilon \approx 1.66 - 1.12i$ and our numerically obtained value of $\ 1.67 - 1.1i$.
We also find good agreement for the stress tensor, labeled $T$, with predicted scaling dimension $\Delta_T = 2$ at $S=0$ and $k=\pm 4\pi/L$, and the $O(N)$ current, labeled $J$, with predicted scaling dimension $\Delta_J = 1$ at $S=1$ and $k=\pm 2\pi/L$ (see Fig.~\ref{fig:spectrum}(d)).

The $\ell$-leg watermelon operators appear in the spectrum as clusters of nearly-degenerate states with different $S$ values, at momentum $k=0$ for even $\ell$ and momentum $k=\pi$ for odd $\ell$ (see Fig.~\ref{fig:spectrum}(c) where the clusters are labeled by A,B,C,D for $\ell=1,2,3,4$).
As explained in the \Ch{SM~\cite{SupplementalMaterials}}, we find a perfect matching between the predicted decomposition of $\ell$-leg watermelon operators into $O(\Ch{N})$ irreducible representations~\cite{10.21468/SciPostPhys.14.5.092} and the clusters of nearly-degenerate spin multiplets we observe in our spectrum. This approximate degeneracy between distinct spin multiplets arises due to the emergent non-invertible symmetry mentioned above~\cite{Jacobsen2023}, and should thus become exact in the thermodynamic limit.

\Ch{\emph{Other lattice realizations of the CCFT.}---Based on our field theory analysis, we expect the CCFT to arise in a broad class of microscopic lattice models with $O(N>2)$ symmetry, and here we provide further numerical evidence for this prediction. Replacing the next-nearest-neighbor hopping $J_2$ with a next-nearest-neighbor biquadratic coupling $K_2$, for instance, yields CCFTs at the points $(K_1, K_2)_{\pm} = (0.385\pm0.243i, -0.414\mp 1.77i)$, as detailed in the End Matter. The CCFT is also not limited to spin-$1$ systems: in the End Matter, we use an analytical mapping between a spin-$1/2$ ladder Hamiltonian and the dilute Temperley-Lieb algebra underlying $O(N)$ loop models~\cite{https://doi.org/10.48550/arxiv.q-alg/9511020,Jacobsen2023,Bellette2023} to identify several \emph{lines} of CCFT points in the ladder model with an exact realization of the non-invertible symmetry discussed earlier~\cite{Jacobsen2023}. These results further support the universality of the CCFT.}

\begin{figure} [t!]
    \centering
    \includegraphics[width=0.98\linewidth]{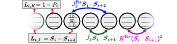}
    \caption{
    {Realization of a CCFT in a spin-1 Heisenberg chain.} Non-Hermiticity can be engineered by introducing jump operators $\boldsymbol S_i -   \boldsymbol S_{i+2}$ and $1-P_0$, where $P_0 = (1/3) [(\boldsymbol{S}_i \cdot \boldsymbol S_{i+1})^2-1]$ is a projector onto the singlet sector. After post-selection on no-click trajectories, the dynamics are governed by a non-Hermitian Hamiltonian with a complex $J_2$ and $K$, which can be tuned to a CCFT.}\label{fig:exp}
\end{figure}

\textit{Realization with monitored quantum dynamics.}--- 
\Ch{Having identified CCFT fixed points in several} simple microscopic non-Hermitian \Ch{Hamiltonians (NHHs), we now turn to the question of whether they can} be realized in quantum dynamics.
\Ch{For concreteness, we focus on the spin-1 $J_1$-$J_2$ model of Eq.~(\ref{eq:NHH}).}
Time evolution with a NHH can be realized by continuous monitoring and post-processing to select only trajectories with no clicks \cite{PhysRevA.106.042210,Ashida02072020,doi:10.1142/S1230161222500044}.
In that case, up to a normalization, the dynamics is described by $\ket{\psi(t)} = e^{- i H t} \ket{\psi(0)}$. Denoting the right eigendecomposition of $H$ as
\bea
H \ket{\psi_{R,\mu}} = (\epsilon_\mu - i \gamma_\mu) \ket{\psi_{R,\mu}},
\eea
we identify from the spectrum frequencies $\epsilon_\mu \in \mathbb{R}$ and decay rates $\gamma_\mu \in \mathbb{R}_+$.

As shown in Fig.~\ref{fig:spectrum}(b), the state corresponding to the CFT vacuum $\ket{\psi_\text{CFT}}$ is not only the ground state (i.e., with lowest energy $\epsilon$), but it is also the ``longest-lived state'' LLS (i.e., with lowest decay rate $\gamma$).
Any initial state with a non-zero overlap with the LLS will asymptotically approach it at late times, and thus we have $\ket{\psi(t \to \infty)} \propto \ket{\psi_\text{CFT}}$.
This shows that dissipation naturally drives the system into the CCFT, and thus provides a way of preparing a highly non-trivial, long-range entangled CFT ground state through engineered dissipation. More generally, $e^{-i H t}$ is guaranteed to relax the system towards a CFT primary state so long as the velocity has negative imaginary part ($\mathrm{Im}(v) < 0$).  Indeed, in that case the primary operator is necessarily the longest-lived state within its conformal family since descendants have decay rates which increase in increments of $(2\pi |\mathrm{Im}(v)|)/L$ (see Fig.~\ref{fig:spectrum}(b)).

We now give an explicit Lindblad master equation whose no-clicks trajectories realizes the non-Hermitian spin-1 Heisenberg chain of Eq.~(\ref{eq:NHH}).
The open system is modeled by a density matrix $\rho$ that obeys the Lindblad master equation
\begin{equation} \label{eq:lindblad}
    \partial_t\rho = -i[H_0,\rho] \Ch{+} \sum_a  (L_{a} \rho L_{a}^{\dagger} - \frac{1}{2} \{ L_{a}^{\dagger} L_{a}, \rho \}),
\end{equation}
where $H_0$ is the Hermitian Hamiltonian of the closed system and $L_{a}$ are the jump operators that describe coupling to the environment. When postselecting for no-click trajectories, the system evolves under an effective non-Hermitian Hamiltonian $H_{\text{eff}} = H_0 - \frac{i}{2} \sum_{a} L^{\dagger}_{a} L_{a}$ \cite{PhysRevA.106.042210,Ashida02072020,doi:10.1142/S1230161222500044}. Here, we choose for the closed system Hamiltonian
\begin{equation}
    H _0= \sum_{i} [J_1 \boldsymbol{S}_{i} \cdot \boldsymbol{S}_{i+1} + J_2^{\mathrm{Re}}\boldsymbol{S}_i \cdot \boldsymbol{S}_{i+2} + K^{\mathrm{Re}} (\boldsymbol{S}_i \cdot \boldsymbol{S}_{i+1})^2],
\end{equation}
and engineer the imaginary terms for $J_2$ and $K$ via the jump operators $\boldsymbol{L}_{i, J} =\sqrt{J_2^{\mathrm{Im}}}( \boldsymbol S_i - \boldsymbol S_{i+2})$ and $L_{i, K} =\sqrt{6K^{\mathrm{Im}}} (1-P_0)$, where $i = 1,...,L$ and $P_0 = (1/3) [(\boldsymbol{S}_i \cdot \boldsymbol S_{i+1})^2-1]$ is the projector onto the singlet sector. The resulting $H_\text{eff}$ realizes the same Hamiltonian we studied numerically (Eq.~(\ref{eq:NHH})) with $J_2 = J^{\mathrm{Re}}_2 + iJ^{\mathrm{Im}}_2$ and $K = K^{\mathrm{Re}} + i K^{\mathrm{Im}}$.
(There is also an additional constant imaginary shift $\delta \equiv -i (4K^{\mathrm{Im}}+ \Ch{2} J^{\mathrm{Im}}_2)L$).

\textit{Discussion}.--- 
We have established the existence of a critical fixed point of the 2D $O(N)$ NLSM for $N>2$ at a complex value of the coupling $g$.
An immediate implication is for perturbation theory: at any fixed loop order, a truncated beta function $\beta(g)$ must generically admit zeros away from the positive real axis.
It would be interesting to assess whether these perturbative zeros align with the CCFT points identified here, providing a nontrivial consistency check on the perturbative expansion~\cite{PhysRevLett.57.1383}\Ch{, see recent discussion in Ref.~\cite{yamamoto2026complexnonlinearsigmamodel}.}
More broadly, whether these complex fixed points could influence the flow on the real axis in the NLSM remains to be determined~\cite{PhysRevE.107.014117}.

Our microscopic analysis so far focused on the fixed point, but it would be valuable to map out the surrounding phase diagram.
Based on the schematic flow in Fig.~\ref{fig:intro}, one expectation is that the critical point manifests as an ``unnecessary transition'' \emph{within} the Haldane phase, in the sense that all trajectories in its vicinity ultimately run to the same strong coupling fixed point.
The situation could be richer, however, as in loop models where a CCFT describes the endpoint of a first-order line separating a ``loop gas'' from a ``loop liquid''~\cite{PhysRevLett.131.131601}.

Finally, we proposed a protocol to prepare the CCFT ground state from generic initial states via continuous monitoring and postselection in a spin-1 Heisenberg chain. Our analysis so far has focused primarily on spectral properties of the corresponding non-Hermitian Hamiltonian; it would be interesting in future work to identify more directly accessible signatures of the CCFT under this dynamics, for example in correlation functions.
Another especially tantalizing direction is to realize this exotic criticality directly at the master-equation level, without the need for postselection. We also note that alternative realizations of complex $O(N)$ nonlinear sigma models may be possible in monitored dynamics of free fermions~\cite{PhysRevX.13.041045}.

\begin{acknowledgments}
We thank Slava Rychkov, Sagar Vijay, Ehud Altman, Kazuaki Takasan, Steven White, and Zlatko Papic for insightful discussions. This work was supported by the U.S. Department of Energy, Office of Science, Office of Basic Energy Sciences under Early Career Research Program Award Number DE-SC0025568. C.Y. gratefully acknowledges support from the Eddleman Quantum Institute postdoctoral fellowship.
\end{acknowledgments}

\bibliography{main.bib}

\onecolumngrid

\newpage
\begin{center}
\ \vskip 0.2cm
{\large\bf End Matter}
\end{center}
\twocolumngrid

\textit{Entanglement entropy and central charge.}---\Ch{We provide further evidence of the CCFT by calculating the central charge via the entanglement entropy.} For non-Hermitian Hamiltonians, the generalized entanglement entropy is a complex quantity \cite{n578-ljd5,PhysRevResearch.2.033069,PhysRevLett.128.010402,Guo_2021,https://doi.org/10.48550/arxiv.2403.03259,PhysRevLett.119.040601,10.21468/SciPostPhys.12.6.194,PhysRevLett.130.241602,10.21468/SciPostPhysCore.6.3.062,PhysRevB.107.205153}. It is defined using the biorthonormal ground-state density matrix $\rho =|\psi_R\rangle\langle \psi_L|$, where $|\psi_R\rangle$ and $\langle \psi_L|$ denote the right and left ground states satisfying $H|\psi_R\rangle = E_g |\psi_R\rangle$ and $\langle \psi_L|H = \langle \psi_L| E_g$. The entanglement entropy $S(l) = \Ch{-}\mathrm{Tr}_A(\rho_A \log \rho_A)$ follows from the reduced density matrix $\rho_A = \mathrm{Tr}_B(\rho)$, obtained by tracing out a subsystem $B$ of length $L - l$. At criticality, $S(l)$ displays the characteristic logarithmic scaling controlled by the central charge $c$, $S(l) = \frac{c}{3} \log \left( \frac{L}{\pi} \sin \frac{\pi l}{L} \right) + s_0,$ where $s_0$ is a non-universal constant and $c$ may be complex \cite{n578-ljd5,10.21468/SciPostPhys.12.6.194,Calabrese_2004,HOLZHEY1994443,PhysRevLett.90.227902,PhysRevLett.104.095701}. The numerical results were performed with DMRG on iTensor using the Arnoldi diagonalization routine \cite{PhysRevLett.69.2863,10.21468/SciPostPhysCodeb.4}, with the left and right states obtained separately by minimizing the real part of the ground-state energy of $H$ and $H^\dagger$ \cite{n578-ljd5}. For a chain of length $L = 14$ with periodic boundary conditions, the optimal fit on the interval $l \in [2,12]$ yields $c = 1.529 -0.161i$ within $2\%$ of the predicted value $c_+ = 1.51 - 0.158i$. \Ch{In the Supplemental Materials (SM)~\cite{SupplementalMaterials}, we extend the DMRG calculation to demonstrate that the central charge estimate changes minimally as a function of larger system sizes $L$.}

{\color{black}

\textit{Universality of the CCFT.}---The mapping of the spin-$1$ model presented in our work to the NLSM implies that the CCFT should appear in a broad class of spin Hamiltonians with $O(3)$ symmetry. As a demonstration of this universality, we identify the same CCFT in two other non-Hermitian spin models: another spin-$1$ Heisenberg chain, and a spin-$1/2$ Heisenberg ladder. In the latter, we identify several \textit{lines} of CCFT points in its parameter space.

\textit{$K_1$-$K_2$ model.}---
Our first example is the spin-$1$ Heisenberg chain with nearest- and next-nearest neighbor biquadratic terms: 
\begin{equation}\label{eq:NHH2}
    H = \sum_{i} \left[ J_1 \boldsymbol{S}_{i} \cdot \boldsymbol{S}_{i+1}+ K_1 (\boldsymbol{S}_i \cdot \boldsymbol{S}_{i+1})^2 + K_2(\boldsymbol{S}_{i} \cdot \boldsymbol{S}_{i+2})^2\right].
\end{equation}
A similar optimization scheme as the one used in the main text (see the SM~\cite{SupplementalMaterials} for details), performed for $L = 12$, identifies the following pair of critical points $(K_1, K_2)_{\pm} = (0.385\pm0.243i, -0.414\mp 1.77i),$ where we have set $J_1 = 1$. The numerically-calculated central charge at $L =12$, $c = 1.565-0.2i$, is within $5\%$ of the predicted value. The scaling dimensions match the predictions closely for low-energy states, such as the $\ell = 1,2$ WMs, see Fig.~\ref{fig:twomodel}(a) and (e), but exhibit poorer agreement with states higher in the spectrum, presumably due to stronger finite size effects in this model compared to the one in the main text.

\begin{figure} [t!]
    \centering
    \includegraphics[width=0.95\linewidth]{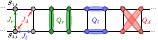}
    \caption{
   \Ch{Illustration of the spin-$1/2$ ladder model realizing the CCFT with an exact non-invertible symmetry [see Eq.~(\ref{eq:ladderh})].}}\label{fig:spinladder}
\end{figure}

\textit{Ladder model with exact non-invertible symmetry.}---
Our second example is a spin-$1/2$ Heisenberg ladder with an exact non-invertible symmetry which prevents loop crossings and thus allows for finding the CCFT numerically by tuning a single complex parameter. The ladder consists of two legs indexed by $\eta = 1,2$, with spin operators $\boldsymbol{S}_{\eta,i}$, where $i$ denotes the rung index \cite{PhysRevB.80.014426,HikiharaStarykh2010}. We begin with a general $\mathrm{SU(2)}$-symmetric ladder Hamiltonian:
\begin{equation} \label{eq:ladderh}
    H = \sum_{i} \sum_{\mu \in \{r,l,d\}} (J_\mu \hat{\mathcal{L}}_i^\mu+ Q_\mu \hat{\mathcal{P}}_i^\mu ).
\end{equation}
Here we include two sets of operators, giving a total of six parameters, see Fig.~\ref{fig:spinladder}. The two-site operators are $\hat{\mathcal{L}}^r_i = \boldsymbol{S}_{1,i} \cdot \boldsymbol{S}_{2,i}$ on the rungs, $\hat{\mathcal{L}}^l_i =\boldsymbol{S}_{1,i} \cdot \boldsymbol{S}_{1,i+1} +\boldsymbol{S}_{2,i} \cdot \boldsymbol{S}_{2,i+1}$ on the legs, and $\hat{\mathcal{L}}^d_i =\boldsymbol{S}_{1,i} \cdot \boldsymbol{S}_{2,i+1} +\boldsymbol{S}_{2,i} \cdot \boldsymbol{S}_{1,i+1}$ on the diagonals. The plaquette operators are $\hat{\mathcal{P}}^{r}_i = (\boldsymbol{S}_{1,i} \cdot \boldsymbol{S}_{2,i})(\boldsymbol{S}_{1,i+1} \cdot \boldsymbol{S}_{2,i+1})$ on the rungs, $\hat{\mathcal{P}}^{l}_i = (\boldsymbol{S}_{1,i} \cdot \boldsymbol{S}_{1,i+1})(\boldsymbol{S}_{2,i} \cdot \boldsymbol{S}_{2,i+1})$ on the legs, and $\hat{\mathcal{P}}^{d}_i = (\boldsymbol{S}_{1,i} \cdot \boldsymbol{S}_{2,i+1})(\boldsymbol{S}_{2,i} \cdot \boldsymbol{S}_{1,i+1})$ on the diagonals.

\begin{figure*} [t!]
    \centering
    \includegraphics[width=1\linewidth]{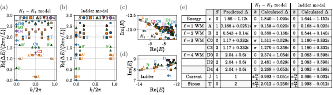}
    \caption{
    \Ch{{Identification of the CCFT in the $K_1$-$K_2$ spin-$1$ chain and the spin-$1/2$ ladder at the Yang-Baxter (YB) point.} (a)-(b) Real part of the extracted scaling dimensions, obtained as rescaled energy gaps $\Delta E / (2 \pi v / L)$, where $\Delta E = E - E_g$ and $E_g$ is the ground state energy and $v$ is the complex velocity, given by $v =0.578-0.72i$ for the $K_1$-$K_2$ model and $v = 0.767+0.235i$ for the ladder model. (c)-(d) Energy spectrum in the complex plane without rescaling by $2\pi v / L$, with the dashed line parameterized by $E_g + vx$, where $x \in \mathbb{R}^+$ showing the direction of the conformal towers. (e) Table of identified CCFT states, comparing the predicted and numerically calculated scaling dimensions $\Delta$. For the ladder model, note the exact degeneracy between different multiplets corresponding to the same $\ell$-leg WM operator thanks to the exact non-invertible symmetry.}}\label{fig:twomodel}
\end{figure*}

The crucial insight is that there exists a five-dimensional subspace of parameters in Eq.~(\ref{eq:ladderh}), defined by the constraint $Q_l + Q_d + 4(J_l + J_d) = 0$, for which the Hamiltonian maps exactly onto the dilute Temperley-Lieb (dTL) algebra underlying the $O(N=3)$ loop model \cite{https://doi.org/10.48550/arxiv.q-alg/9511020,Jacobsen2023,Bellette2023}. 
We introduce a convenient parametrization of this five-dimensional space: $H = \sum_i h_i$ with 
\begin{equation}
h_i = J_r \hat{\mathcal{L}}_{i}^{\,r} + Q_r \hat{\mathcal{P}}_{i}^{\,r} + J_- \hat{\mathcal{L}}_{i}^{\,-} + Q_- \hat{\mathcal{P}}_{i}^{\,-} + \mathcal{V}_+ \bigl(4\hat{\mathcal{P}}_{i}^{\,+}-\hat{\mathcal{L}}_{i}^{\,+}\bigr)
\end{equation}
with
$
    \hat{\mathcal{L}}_{i}^{\pm} = \hat{\mathcal{L}}^l_i \pm \hat{\mathcal{L}}^d_i$ and $\hat{\mathcal{P}}_{i}^{\pm} = \hat{\mathcal{P}}^l_i \pm \hat{\mathcal{P}}^d_i
$.
(The mapping between spins and loops proceeds as follows: after decomposing the spin Hilbert space on each rung into spin-$0$ ($\ket{0}$) and spin-$1$ states (spanned by $\{\ket{x},\ket{y},\ket{z}\}$), one identifies these, respectively, with a vacant site or with a site occupied by a loop strand. One can then map loop-connectivity configurations to kets in the spin Hilbert space, such that two occupied sites connected by a loop form a singlet state given by $\frac1{\sqrt{3}} \sum_{a \in x,y,z} \ket{aa} $. For example, this mapping gives $\frac1{3}\sum_{a,b \in x,y,z}\ket{a b 0 b 0 a } \leftrightarrow \lvert\smash{\tikz[scale=.58,baseline=-.5ex]{
  \foreach \x in {0,.26,.52,.78,1.04,1.30}
    \fill (\x,.24) circle (1.05pt);
  \draw[line width=.5pt,line cap=round]
    (.26,.24) .. controls (.39,-.02) and (.65,-.02) .. (.78,.24);
  \draw[line width=.5pt,line cap=round]
    (0,.24) .. controls (.32,-.34) and (.98,-.34) .. (1.30,.24);
}}\rangle$.)

Since this entire five-dimensional family of Hamiltonians lies within the dTL algebra, it possesses an exact non-invertible symmetry, as explained in Ref.~\cite{Jacobsen2023}. As a result, these Hamiltonians generate only loop-crossing-free configurations (in spin language, the long-range singlets connecting different rungs do not cross). This model therefore does not suffer from the same practical difficulty as in the main text, where a second parameter had to be tuned to cancel weakly irrelevant loop crossings and obtain good finite-size convergence.
We should therefore be able to find the CCFT by tuning a single (complex) parameter within this five-parameter space, as we confirm numerically below.

As a representative example, we choose the pair of parameters $Q_r$ and $J_-$ from the five-dimensional space and study how the CCFT moves in the complex plane $(\mathrm{Re}(J_-), \mathrm{Im}(J_-))$ as $\mathrm{Re}(Q_r)$ is varied. In this way, we trace out a line of CCFTs; see Fig.~\ref{fig:critcurve}. To determine the critical curve, we vary the real part of $Q_r$ and optimize over the complex coefficient $J_{-}$ to locate the finite-size critical point, defined as the point that minimizes the residual
\begin{equation} \label{eq:residual}
    \mathcal{F}(L_1,L_2) = |L_1 \Delta E(L_1) - L_2 \Delta E(L_2)|,
\end{equation}
where $L_1$ and $L_2$ are two chain lengths, and $\Delta E = E_{S=1} - E_{g}$ is the ``triplet gap'' separating the complex energy of the ground state from that of the lowest-real-part state in the triplet sector. (At the CCFT, this is the gap corresponding to the $l=1$ WM operator).
This residual quantifies how closely the spectrum follows the expected $1/L$ scaling and should therefore be minimized at a critical point. Fig.~\ref{fig:critcurve}(a) shows the resulting critical curve. 
In Panel (b), we show a representative heatmap of the residual in the complex plane of $J_{-}$, exhibiting a local minimum.
 Panels (c)–(e) give the extracted central charge and scaling dimensions along the critical curve, which show close agreement with the predicted values.

\begin{figure} [h!]
    \centering
    \includegraphics[width=0.98\linewidth]{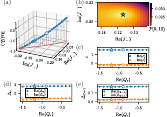}
    \caption{
     \Ch{(a) Curve of CCFT points obtained by varying the coupling $\mathrm{Re}[Q_r]$ and optimizing over $J_-$ while keeping all other parameters fixed at the value of the Yang-Baxter (YB) point. The open circle marks the YB point. (b) Residual (Eq.~(\ref{eq:residual})) showing a minimum at the finite-size location of the critical point, marked by the star. (c)-(e) Central charge and scaling dimensions of $\epsilon$ and $\ell=1$ WM along the critical curve, computed at $L = 10$, showing close agreement with the predicted values (dashed lines).}}\label{fig:critcurve}
\end{figure}

To reduce the numerical cost of searching for the CCFT in this large parameter space, we use as a ``seed'' for the CCFT line a point  analytically predicted to realize the CCFT. We refer to this point as the YB point (shown as open circle in Fig.~\ref{fig:critcurve}) with coordinates $J_- \approx -0.0450-0.249i$, $Q_r \approx -0.894$, $J_r \approx 0.447+0.894i$, $Q_- \approx -3.398+0.994i$, $\mathcal{V}_+ \approx 0.224-0.224i$ (see the SM~\cite{SupplementalMaterials} for analytical expressions). It was obtained from a known solution of the Yang-Baxter (YB) equation, see Ref.~\cite{https://doi.org/10.48550/arxiv.q-alg/9511020}, analytically continued from $N\leq 2$ to $N>2$ in a manner similar to Ref.~\cite{PhysRevLett.131.131601}. The spectrum at the YB point, shown in Fig.~\ref{fig:twomodel}(b), (d), and (e) for $L = 10$, is in very good agreement with the CCFT predictions. 

Finally, using the same procedure, we obtained two additional CCFT curves passing through the YB point along different directions for two other choices of parameter pairs; see the SM~\cite{SupplementalMaterials}.
}

\end{document}


\widetext
\onecolumngrid

\begin{center}
\textbf{\large Supplemental Material:
\\Asymptotic freedom, lost: Complex conformal field theory in the two-dimensional $O(N>2)$ nonlinear sigma model and its realization in Heisenberg spin chains}
\\[0.4ex] Christopher Yang and Thomas Scaffidi
\end{center}
\par
\setcounter{page}{1}
\twocolumngrid

\setcounter{equation}{0}
\setcounter{figure}{0}
\setcounter{table}{0}
\setcounter{page}{1}
\makeatletter
\renewcommand{\theequation}{S\arabic{equation}}
\renewcommand{\thefigure}{S\arabic{figure}}
\setcounter{secnumdepth}{4}

\section{Identification of watermelon operators in the spectrum}
In this section, we discuss the identification of $\ell$-leg watermelon operators in the CCFT spectra [see Fig.~3 in the main text and Fig.~6 in the End Matter] for $\ell=1,2,3,4$. For each $\ell$ value, we find the decomposition into $O(N)$ irreducible representations in Ref.~\cite{10.21468/SciPostPhys.14.5.092}, and we then specialize to $N=3$ (for a reference on the multiplicity of Young tableaux of $O(N)$, see \cite{Cvitanovic2008}). We find perfect agreement with the multiplicity we observe in our numerics [see clusters labeled A, B, C, D in Fig.~3(c) in the main text].
For $\ell=1$, Ref.~\cite{10.21468/SciPostPhys.14.5.092} gives the $O(N)$ irreducible representation $[1]$ (using row-length notation for Young tableaux), i.e. an $S=1$ state in the case of $N=3$. This is the $O(N)$ vector field. We confirm this numerically (point A).
For $\ell=2$, Ref.~\cite{10.21468/SciPostPhys.14.5.092} gives $[2]$, i.e. an $S=2$ state, which we confirm numerically (point B). This is the symmetric traceless rank-2 tensor.
For $\ell=3$, Ref.~\cite{10.21468/SciPostPhys.14.5.092} gives $[3]+[111]$, which for $N=3$ specializes to $3 \oplus 0$, i.e. a spin-3 and a singlet. Microscopically, the singlet operator is identified with $\boldsymbol S_i \cdot ( \boldsymbol S_j \times \boldsymbol S_k)$. We confirm the presence of both the spin 0 (point C0) and the spin 3 (point C3) states.
For $\ell=4$, Ref.~\cite{10.21468/SciPostPhys.14.5.092} gives $[4]+[2]+[0]+[22]+[211]$. For $O(3)$, the multiplicity of $[22]$ and $[211]$ is zero (see e.g. Table 10.3 and 10.4 of \cite{Cvitanovic2008}), leaving only $4 \oplus 2 \oplus 0$.
The singlet state is the previously discussed loop crossing operator that appears in the action and is weakly irrelevant. We have confirmed the presence of a $S=4$, $S=2$ and $S=0$ state close to the predicted $\ell=4$ WM scaling dimension (see points D0, D2, and D4). We summarize all the calculated scaling dimensions in Fig.~3(d) in the main text.

\section{Optimization Procedure for the Spin-$1$ Heisenberg Chain} \label{sec:opproc}
In this section, we describe the optimization algorithm used to find the CCFT point in the spin-$1$ Heisenberg models. In particular, we use a gradient descent algorithm to locate the finite-size critical point in the parameter space $(J_2,K)$ or $(K_1, K_2)$ with the fastest finite-size convergence. We do so by matching the thermal and $\ell = 1$ watermelon scaling dimensions in the spin chain spectrum with their predicted values. To extract these scaling dimensions, we compare the momentum $k$ and total spin $S$ of the eigenstates to the known symmetries of the CCFT primaries, see Refs.~\cite{10.21468/SciPostPhys.14.5.092,Gorbenko2020}. We will use $E_{k,S}^\nu$ to denote the eigenergies and $\nu \in \mathbb{Z}^+$ to index the states in order of increasing real energy. The CCFT vacuum is a translationally-invariant singlet and naturally corresponds to the Haldane ground state with energy $E_g = E_{k=0,S=0}^0$.
The thermal operator is a singlet, so it corresponds to $E_{k=0,S=0}^1$. The $\ell=1$ WM is the $O(3)$ vector operator and has momentum $k=\pi$ since this is an antiferromagnetic model: it thus corresponds to $E_{k=\pi,S=1}^0$. Finally, the velocity 
\begin{equation} \label{eq:velocity}
    v = \frac{L}{2\pi} \big(E_{k=\pi+\frac{2\pi}{L},S=1}^0 - E_{k=\pi,S=1}^0\big)
\end{equation}
can be inferred from the first descendant state of the $\ell = 1$ WM at $k=\pi + 2\pi/L$. The resulting $\ell = 1$ WM and thermal scaling dimensions are respectively given by
\begin{equation}
    \Delta_{\ell=1}^{\text{calc}} = 
    \frac{E_{k=\pi,S=1}^0-E_g}{2\pi v/L}, \text{ and } 
    \Delta_\epsilon^{\text{calc}} = 
    \frac{E_{k=0,S=0}^1-E_g}{2\pi v/L}.
\end{equation}

Using the calculated scaling dimensions from exact diagonalization, we can now locate the finite-size critical point with fastest convergence by searching for the minima of the cost function $f(J_2, K)=|R_1| + w|R_2|$ through gradient descent, where 
\begin{equation}
    R_1=\frac{\Delta_{\ell=1}}{\Delta_\epsilon}-\frac{\Delta_{\ell=1}^{\text{calc}}}{\Delta_\epsilon^{\text{calc}}}, 
    \text{ }
    R_2=\frac{\Delta_\epsilon}{\Delta_\epsilon^{\text{calc}}}-1.
\end{equation}
and $w = 0.2$ is a feature scaling weight which we found to help the gradient descent algorithm. We emphasize that the two-parameter optimization over $J_2$ and $K$ or $K_1$ and $K_2$ is used solely to improve finite-size numerics, and only a single complex tuning parameter should be needed to realize the CCFT in the thermodynamic limit.

\section{Central Charge of the Spin-$1$ Heisenberg Chain at Larger System Sizes}
\begin{figure} [t!]
    \centering
    \includegraphics[width=0.98\linewidth]{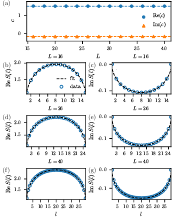}
    \caption{
     Central charge vs. system size $L$ as calculated via DMRG (with dashed lines indicating the predicted values), and the real and imaginary parts of the complex entanglement entropy $S(l)$ for different subsystem sizes $l$ at the CCFT.}\label{fig:clargel}
\end{figure}

To verify that the finite-size fixed point identified in our work lies close to the true fixed point, we use DMRG to calculate the central charge at larger system sizes. Here, we focus on the spin-$1$ Heisenberg chain with parameters $(J_2, K)$ set to the identified finite-size fixed point, see Eq.~(6) in the main text. In Fig.~\ref{fig:clargel}(a), we show that the resulting central-charge calculation remains approximately independent of system size, with excellent agreement with the expected logarithmic entanglement scaling, as illustrated in panels (b)–(g).

\section{Ground State Energy Scaling of the Spin-$1$ Heisenberg Chain}
\begin{figure} [t!]
    \centering
    \includegraphics[width=\linewidth]{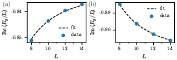}
    \caption{
     Real and imaginary parts of the ground-state energy density $E_g / L$ and fit to $E_g / L = e_\infty - \pi v c / (6 L^2)$, where $e_\infty$ is a fitting parameter, $c$ is determined from the entanglement entropy, and $v$ is obtained from the energy spectrum. 
     }\label{fig:sm1}
\end{figure}

As an additional consistency check for the finite-size critical point in the $(J_2, K)$-model, we verify the finite-size scaling of the ground state energy
\begin{equation}
E_g = e_{\infty} L - \frac{\pi v c}{6 L} + O(L^{-2}),
\end{equation}
governed by the central charge $c$, where $e_{\infty}$ denotes the non-universal bulk energy density and $v$ is the velocity \cite{PhysRevLett.56.742,DiFrancesco1997,https://doi.org/10.48550/arxiv.0807.3472}.
In Fig.~\ref{fig:sm1}, we plot $E_g / L$ and fit the data using $e_{\infty}$ as a free parameter. Here, we have used the value of $c$ obtained from the entanglement entropy and the velocity determined from the spectrum (see Sec.~\ref{sec:opproc} for details). The agreement indicates that the finite-size scaling is consistent with the CCFT prediction.

\section{CCFT lines in a spin-1/2 ladder Hamiltonian and Mapping to the Dilute-TL algebra}

In the End Matter, we presented two additional non-Hermitian lattice realizations of the CCFT. In this section, we provide details of the spin-$1/2$ ladder and its mapping to the dilute Temperley-Lieb (dTL) algebra describing the  classical $O(N)$ loop models.

\subsection{Dilute TL Algebra} \label{sec:dtlalg}
The dTL algebra for loop fugacity $N$ can be defined on a $1$D chain with an $N+1$-dimensional Hilbert space on each site,  spanned by $\{\ket{v},\ket{1},\ket{2},\dots,\ket{N}\}$. The state $\ket{v}$ corresponds to a vacant site, whereas the states $\ket{a}$ with $a = 1,2,...,N$ correspond to sites occupied by a loop strand, where $a$ runs over the components of the classical $O(N)$ vector underlying the loop model~\cite{https://doi.org/10.48550/arxiv.q-alg/9511020,Jacobsen2023}.
We will focus on $N = 3$, for which the states corresponding to $a=1,2,3$ are realized respectively with the $\ket{x},\ket{y},$ and $\ket{z}$ states of a spin-$1$:
\begin{align}
|x\rangle &\equiv \frac{i}{\sqrt{2}}\left(|S^z=+1\rangle - |S^z=-1\rangle\right), \\
|y\rangle &\equiv \frac{1}{\sqrt{2}}\left(|S^z=+1\rangle + |S^z=-1\rangle\right), \\
|z\rangle &\equiv i|S^z=0\rangle .
\end{align}

The dTL algebra is generated by a set of $9$ operators~\cite{https://doi.org/10.48550/arxiv.q-alg/9511020}. It contains four projectors on sites $i$ and $i+1$, 
\begin{equation}
\begin{split}
     (\rangle \langle)_i &= n_i n_{i+1}  \\
     (\rangle \ )_i &= n_i (1- n_{i+1})\\
     (\ \langle)_i &= (1-n_i) n_{i+1}   \\
     (\ \ )_i &= (1-n_i)(1-n_{i+1})
\end{split}
\end{equation}
onto the $(o,o)$, ($o,v)$, $(v,o)$, and $(v,v)$ sectors on the two spin sites, respectively, where $o$ is shorthand for an occupied state, $n_i = 1-P_i^v$ is the occupancy on site $i$, and $P^v = \ket{v}\bra{v}$ is the single-site projector onto the vacant sector. It also contains ``hopping'' terms 
\begin{equation}
\begin{split}
    (\backslash)_i &= \sum_{a=x,y,z} |a,v\rangle \langle v,a|, \\
     (/)_i &= \sum_{a=x,y,z} |v,a\rangle \langle a,v|
\end{split}
\end{equation}
which exchanges a vacancy with an occupancy. Lastly, it also contains singlet projection operators
\begin{equation}
\begin{split}
    (\spacecap)_i &= \sqrt{3} |\Omega \rangle \langle v,v| \\
    (\cupspace)_i &= \sqrt{3}  |v,v\rangle \langle \Omega| \\
    e_i \equiv (\cupcap)_i &= 3 |\Omega \rangle \langle \Omega|,
\end{split}
\end{equation}
where 
\begin{equation}
    |\Omega\rangle = \frac{1}{\sqrt{N}} \sum_{a=x,y,z } |a,a\rangle 
\end{equation}
is the singlet state. (In the $\ket{S^z_i,S^z_{i+1}}$ basis, it reads $\ket{\Omega} = 3^{-1/2}(\ket{1,-1} - \ket{0,0} + \ket{-1,1})$.)

Under inversion and translation symmetry, the most general Hamiltonian on the dTL space is written as, $H = \sum_i h_i$, where
\begin{equation} \label{eq:hdtl}
    h_i = J e_i + t[(\backslash)_i + (/)_i] + g [(\spacecap)_i + (\cupspace)_i] + h n_i + V n_i n_{i+1} .
\end{equation}

\subsection{Critical Yang-Baxter Hamiltonian}
A special point in the 5-dimensional parameter space of Eq.~\ref{eq:hdtl} can be identified that corresponds to the analytical continuation to $N>2$ of well-known lines of fixed points occurring at $N \leq 2$ in loop models. (This is related to the critical points studied for transfer matrices of loop models in Ref.~\cite{PhysRevLett.131.131601}).
We call this point the YB point, since it solves the Yang-Baxter equation. To find the Hamiltonian, we use the YB face operator $X_i(u)$ derived in Ref.~\cite{https://doi.org/10.48550/arxiv.q-alg/9511020}. Here, $u$ denotes the Yang-Baxter spectral parameter. In particular, the Hamiltonian is obtained from the first derivative $H = X_i'(u)|_{u=0}$, from which we obtain the coordinates of the YB point:
\begin{equation} \label{eq:critj}
    J = \frac{\sin \lambda}{\sin(2\lambda)\sin(3\lambda)},
\end{equation}
\begin{equation}
    t = -\frac{1}{\sin(2\lambda)},
\end{equation}
\begin{equation}
    g = -\frac{1}{\sin(3\lambda)},
\end{equation}
\begin{equation}
    h = \frac{2\left[2 \cos^2(\lambda)\sin(3\lambda) - \sin\lambda\right]}{\sin(2\lambda)\sin(3\lambda)},
\end{equation}
and
\begin{equation} \label{eq:critv}
    V = -\frac{\cos(2\lambda)}{\cos\lambda\,\sin(3\lambda)}.
\end{equation}
Here, we have $\lambda = \frac{1}{4} \cos^{-1}(-N/2)$ which gives, for $N=3$, $\lambda = \pi/4 - i \ln \phi /2$, where $\phi = (1 + \sqrt{5})/2$.

\subsection{Mapping to a Spin-$1/2$ Ladder}

As mentioned above, for $N=3$, the on-site Hilbert space of the dTL algebra is spanned by $\{\ket{v},\ket{x},\ket{y},\ket{z}\}$, where $\ket{x},\ket{y},\ket{z}$ form the basis of a spin-1.
Identifying $\ket{v}$ as a spin-0 object, it means we have a spin-0 and a spin-1 on each site, and it is thus natural to realize this with a spin-1/2 ladder where we identify each site of the dTL algebra with a rung of the ladder.

The ladder is constructed from two legs, with spin-1/2 operators denoted $\boldsymbol{S}_{1,i}$ and $\boldsymbol{S}_{2,i}$, on the top and bottom legs, respectively, and $i$ denoting the rung index \cite{PhysRevB.80.014426}. The Hamiltonian we consider is given by
\begin{equation} \label{eq:ladderh}
    H = \sum_{i,\mu} (J_\mu \hat{\mathcal{L}}_i^\mu+ Q_\mu \hat{\mathcal{P}}_i^\mu ).
\end{equation}
See Fig 5 of the End Matter for a sketch of the ladder and the six different terms in the Hamiltonian. There are three bond operators:
\begin{equation}
    \hat{\mathcal{L}}^r_i = \boldsymbol{S}_{1,i} \cdot \boldsymbol{S}_{2,i}
\end{equation}
acting on the rung,
\begin{equation}
    \hat{\mathcal{L}}^l_i =\boldsymbol{S}_{1,i} \cdot \boldsymbol{S}_{1,i+1} +\boldsymbol{S}_{2,i} \cdot \boldsymbol{S}_{2,i+1}
\end{equation}
acting on the legs, and 
\begin{equation}
    \hat{\mathcal{L}}^d_i =\boldsymbol{S}_{1,i} \cdot \boldsymbol{S}_{2,i+1} +\boldsymbol{S}_{2,i} \cdot \boldsymbol{S}_{1,i+1}
\end{equation}
acting on the diagonals. The plaquette operators are given by 
\begin{equation}
    \hat{\mathcal{P}}^{r}_i = (\boldsymbol{S}_{1,i} \cdot \boldsymbol{S}_{2,i})(\boldsymbol{S}_{1,i+1} \cdot \boldsymbol{S}_{2,i+1})
\end{equation}
acting on the rungs,
\begin{equation}
    \hat{\mathcal{P}}^{l}_i = (\boldsymbol{S}_{1,i} \cdot \boldsymbol{S}_{1,i+1})(\boldsymbol{S}_{2,i} \cdot \boldsymbol{S}_{2,i+1})
\end{equation}
acting on the legs, 
\begin{equation}
    \hat{\mathcal{P}}^{d}_i = (\boldsymbol{S}_{1,i} \cdot \boldsymbol{S}_{2,i+1})(\boldsymbol{S}_{2,i} \cdot \boldsymbol{S}_{1,i+1}),
\end{equation}
acting on the diagonals.

Each rung of the spin-$1/2$ chain can be interpreted as forming an emergent vacancy (spin-$0$) or occupancy (spin-$1$) on the dTL chain. To make this mapping explicit, we define the emergent spin operator on rung $i$,
\begin{equation}
    \boldsymbol{J}_i = \boldsymbol{S}_{1,i} + \boldsymbol{S}_{2,i}.
\end{equation}
The operator $e_i$ projects onto the singlet sector of two emergent spin-$1$ rungs, so
\begin{equation}
    e_i = 3(P^{(1)}_i P^{(1)}_{i+1}) \Pi_{i,i+1}^{(0)} (P^{(1)}_i P^{(1)}_{i+1}),
\end{equation}
where $\Pi_{i,i+1}^{(0)} =\frac{1}{3} [(\boldsymbol{J}_i \cdot \boldsymbol{J}_{i+1})^2 - 1]$ is the singlet projector on rungs $i$ and $i+1$, and the triplet projector $P_i^{(1)} = 3/4 + \boldsymbol{S}_{1,i} \cdot \boldsymbol{S}_{2,i}$ conditions the operation on rungs $i$ and $i +1$ being occupied (i.e. in spin-$1$ states).

\begin{figure}[t!]
    \centering
    \includegraphics[width=\linewidth]{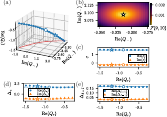}
    \caption{(a) Curve of CCFT points obtained by varying $\mathrm{Re}[Q_r]$ and tuning $Q_-$ to remain critical. The open circle marks the Yang-Baxter (YB) CCFT point located at $Q_r \approx  -0.894$ and $Q_- \approx -3.40+0.994i$. The other parameters are kept constant at their value at the YB point. (b) The critical curve is determined by minimizing the residual, Eq.~(\ref{eq:residual}), which quantifies deviations from the expected $1/L$ scaling. Shown here is the residual in the vicinity of the local minimum marked by the star, for the point indicated on panel (a). (c)-(e) Central charge and scaling dimensions along the critical curve, showing close agreement with the predicted values (dashed lines).}\label{fig:vary1}
\end{figure}

We now find explicit expressions for the hopping operators $(\backslash)_i$ and $(/)_i$ and also for the operators $(\spacecap)_i$ and $(\cupspace)_i$.
These operators can generate transitions between vacant and occupied states on a given rung, which are respectively odd or even under the permutation of two spins on the same rung. The operator that connects these sectors is also necessarily antisymmetric, and the simplest choice is $\boldsymbol{\delta}_i = \boldsymbol{S}_{1,i} - \boldsymbol{S}_{2,i}$. In fact, an explicit calculation shows that
\begin{equation}
    |\Omega\rangle = - \frac{1}{\sqrt{3}} (\boldsymbol{\delta}_i \cdot \boldsymbol{\delta}_{i+1} )|v,v\rangle,
\end{equation}
and
\begin{equation}
    |v,m\rangle = \boldsymbol{\delta}_i \cdot \boldsymbol{\delta}_{i+1} |m,v\rangle.
\end{equation}
These properties provide a simple mapping of the following dTL operators
\begin{equation} \label{eq:cupproj}
    (\cupspace)_i = - P^{(1)}_i P^{(1)}_{i+1} (\boldsymbol{\delta}_{i} \cdot \boldsymbol{\delta}_{i+1}) P^{(0)}_i P^{(0)}_{i+1}
\end{equation}
\begin{equation}
    (\spacecap)_i = - P^{(0)}_i P^{(0)}_{i+1} (\boldsymbol{\delta}_{i} \cdot \boldsymbol{\delta}_{i+1}) P^{(1)}_i P^{(1)}_{i+1}.
\end{equation}
\begin{equation}
    (\backslash)_i = P^{(1)}_i P^{(0)}_{i+1} (\boldsymbol{\delta}_i \cdot \boldsymbol{\delta}_{i+1} ) P^{(0)}_i P^{(1)}_{i+1}.
\end{equation}
\begin{equation} \label{eq:slashproj}
    (/)_i = P^{(0)}_i P^{(1)}_{i+1} (\boldsymbol{\delta}_i \cdot \boldsymbol{\delta}_{i+1} ) P^{(1)}_i P^{(0)}_{i+1}.
\end{equation}
where $P_i^{(0)} = 1/4 - \boldsymbol{S}_{1,i} \cdot \boldsymbol{S}_{2,i}$ is singlet projector on rung $i$. The final operator appearing in the Hamiltonian, see Eq~(\ref{eq:hdtl}), is the density operator $n_i$. Its mapping is straightforward, $n_i = P_i^{(1)}$, because it is simply the projector onto the occupied triplet sector of the rung.

The final step is to express the dTL operators in terms of the plaquette and leg operators of Eq.~\ref{eq:ladderh}. This can be achieved via the following identities on the emergent spin operator $\boldsymbol{J}_i$,
\begin{equation}
    \boldsymbol{J}_i^2 = \frac{3}{2} + 2 \hat{\mathcal{L}}^r_i, \quad \boldsymbol{J}_i \cdot \boldsymbol{J}_{i+1} = \hat{\mathcal{L}}^l_i + \hat{\mathcal{L}}^d_i
\end{equation}
and
\begin{equation}
    (\boldsymbol{J}_i \cdot \boldsymbol{J}_{i+1})^2 = \frac{3}{4} + \hat{\mathcal{L}}^r_i + \hat{\mathcal{L}}^r_{i+1} - \frac{1}{2} (\hat{\mathcal{L}}^l_i + \hat{\mathcal{L}}^d_i) + 2 (\hat{\mathcal{P}}^l_i + \hat{\mathcal{P}}^d_i).
\end{equation}
One can also derive the following identities for the antisymmetric operator $\boldsymbol{\delta}_i$,
\begin{equation}
    \boldsymbol{\delta}_i^2 = \frac{3}{2} - 2 \hat{\mathcal{L}}^r_i \quad \text{and} \quad \boldsymbol{\delta}_i \cdot \boldsymbol{\delta}_{i+1} = \hat{\mathcal{L}}^l_i - \hat{\mathcal{L}}^d_i,
\end{equation}
which relate the operators presented in Eqs.~(\ref{eq:cupproj})-(\ref{eq:slashproj}) to the ladder and plaquette operators. Finally, we complete the mapping by identifying $n_i = \frac{1}{2} \boldsymbol{J}_i^2$ and $n_i n_{i+1} = \frac{9}{16} + \frac{3}{4}(\hat{\mathcal{L}}^r_i + \hat{\mathcal{L}}^r_{i+1}) + \hat{\mathcal{L}}^r_i \hat{\mathcal{L}}^r_{i+1}$.

\begin{figure}[t!]
    \centering
    \includegraphics[width=\linewidth]{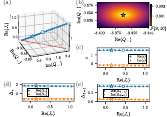}
    \caption{Same as Fig.~\ref{fig:vary1}, but by varying the parameter $\mathrm{Re}[J_r]$ and tuning $Q_-$ to remain critical. The open circle marks the YB point located at $J_r \approx  0.447+0.894i$ and $Q_- \approx -3.40+0.994i$. The other parameters are kept constant at their value at the YB point.}\label{fig:vary2}
\end{figure}

These results provide a linear map $y = M x$ between the parameters of the dTL Hamiltonian [Eq.~(\ref{eq:hdtl})],
\begin{equation}
    x = \begin{pmatrix}
        J, t, g, h, V
    \end{pmatrix}
\end{equation}
and the plaquette and ladder operator coefficients of the spin Hamiltonian [Eq.~(\ref{eq:ladderh})],
\begin{equation}
     y = \begin{pmatrix}
         J_r & J_l & J_d & Q_r & Q_l & Q_d 
     \end{pmatrix}.
\end{equation}
Here,
\begin{equation}
    M = \begin{pmatrix}
        \frac{1}{2} & 0 & 0 & 1 & \frac{3}{2}  \\
        -\frac{1}{2} & \frac{1}{2} & -\frac{1}{2}  & 0 & 0 \\
        -\frac{1}{2} & - \frac{1}{2} & \frac{1}{2} & 0 & 0 \\
        -1 & 0 & 0 & 0 & 1 \\
        2 & 2 & 2 & 0  & 0 \\
        2 & -2 & -2 & 0  & 0 
        
    \end{pmatrix}
\end{equation}
is a rank-$5$ matrix, which implies the presence of a $5$-dimensional hypersurface of dTL Hamiltonians in the $6$-parameter $y$-space of spin-1/2 ladder Hamiltonians.

Since the dTL algebra underlies loop models without crossings---it is easy to see that the generators of the dTL algebra, see Sec.~\ref{sec:dtlalg}, cannot generate crossings---we also call this 5-dimensional space the space of ``loop-crossing-free'' Hamiltonians.
We utilize this loop-crossing-free property in the main text to facilitate the numerical search for CCFT fixed points.
We also note that this loop-crossing-free property was associated with the presence of a non-invertible symmetry operator which commutes with all the dTL generators~\cite{Jacobsen2023}. This non-invertible symmetry is thus present over this entire 5-dimensional space, and its generator could be mapped explicitly to the spin-1/2 ladder using the mapping presented here.
We have confirmed this numerically in the End Matter, where we saw exact degeneracies between distinct $O(3)$ multiplets corresponding to the same $\ell$-leg watermelon operators.

\subsection{YB point in terms of bond and plaquette terms}
Using the mapping above, we can obtain the location of the YB point in terms of the bond and plaquette operators.
Using  Eqs.~(\ref{eq:critj})-(\ref{eq:critv}), we find that
\begin{equation}
    J_r = 2[\cot 2\lambda + (\csc 2 \lambda)/4], 
\end{equation}
\begin{equation}
\ J_l = (1-2\cos \lambda) / [2 \sin(3\lambda)], 
\end{equation}
\begin{equation}
    J_d = (2 \cos \lambda + 1)(\cos \lambda - 1) / [2 \cos \lambda \sin (3\lambda )],
\end{equation}
\begin{equation}
    Q_r = -\csc 2\lambda,
\end{equation}
\begin{equation}
        Q_{l} = -2[\tan\lambda + \cot (3\lambda /2)],
\end{equation}
and
\begin{equation}
    Q_{d} = 2[\cot\lambda + \tan(3\lambda /2)].
\end{equation}
As before,  $\lambda = \frac{1}{4} \cos^{-1}(-N/2) = \pi/4 - i \ln \phi /2$, where $\phi = (1 + \sqrt{5})/2$ for $N = 3$.

\subsection{Critical Curves}
A simple basis for the loop-crossing-free subspace of the spin ladder, i.e., the column space of the matrix $M$, is obtained by writing the Hamiltonian as a linear combination of five operators:
\begin{equation}
h_i = J_r \hat{\mathcal{L}}_{i}^{\,r} + Q_r \hat{\mathcal{P}}_{i}^{\,r} + J_- \hat{\mathcal{L}}_{i}^{\,-} + Q_- \hat{\mathcal{P}}_{i}^{\,-} + \mathcal{V}_+ \bigl(4\hat{\mathcal{P}}_{i}^{\,+}-\hat{\mathcal{L}}_{i}^{\,+}\bigr)
\end{equation}
with
\begin{equation}
    \hat{\mathcal{L}}_{i}^{\pm} = \hat{\mathcal{L}}^l_i \pm \hat{\mathcal{L}}^d_i, \qquad \hat{\mathcal{P}}_{i}^{\pm} = \hat{\mathcal{P}}^l_i \pm \hat{\mathcal{P}}^d_i.
\end{equation}

As explained in the End Matter, we can find lines in this 5-dimensional parameter space that realize the CCFT by varying one parameter, and tuning another (complex) parameter to stay on criticality.
In the End Matter, we presented one such critical line, obtained by varying $Q_r$ and optimizing over $(\mathrm{Re}[J_-],\mathrm{Im}[J_-])$ to minimize the residual
\begin{equation} \label{eq:residual}
    \mathcal{F}(L_1,L_2) = |L_1 \Delta E(L_1) - L_2 \Delta E(L_2)|,
\end{equation}
where $L_1$ and $L_2$ are two chain lengths, and $\Delta E(L)$ is the triplet gap at system size $L$. In Figs.~\ref{fig:vary1} and \ref{fig:vary2}, we present CCFT lines obtained by varying two other pairs of parameters.

\begin{figure} [t!]
    \centering
    \includegraphics[width=0.98\linewidth]{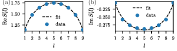}
    \caption{
     Real and imaginary parts of the complex entanglement entropy $S(l)$ for different subsystem sizes at the YB point in the spin-$1/2$ ladder model. The dashed line indicates the fit to the logarithmic scaling expected at criticality, which estimates the central charge to be $c \approx 1.54-0.214i$.}\label{fig:eeladder}
\end{figure}

\subsection{Entanglement Entropy at the YB Point}
Finally, in Fig.~\ref{fig:eeladder}, we show the biorthogonal complex entanglement entropy (see End Matter for definition) of the ladder model at the YB point for $L = 10$, where the cut is taken vertically across legs, and $l$ corresponds to the number of rungs in the subsystem over which the partial trace is taken. The fitted central charge $c \approx 1.54-0.214i$ is in fair agreement with the predicted value $c_+ = 1.51 - 0.158i$.

\bibliographystyle{apsrev4-1}
\bibliography{supplement}